\documentclass{aa}
\usepackage{graphicx}
\usepackage{natbib}  
\usepackage{pdflscape}
\usepackage{txfonts}

\begin{document}

   \title{Stellar population astrophysics (SPA)
with the TNG\thanks{Based on observations made with the Italian Telescopio Nazionale Galileo (TNG) 
operated on the island of La Palma by the Fundación Galileo Galilei of the INAF (Istituto Nazionale di Astrofisica) at the Spanish Observatorio 
del Roque de los Muchachos of the Instituto de Astrofisica de Canarias.
This study is part of the 
Large Program titled {\it SPA - Stellar Population Astrophysics:  the detailed, age-resolved
chemistry of the Milky Way disk} (PI: L. Origlia), granted observing time with HARPS-N and GIANO-B echelle spectrographs
at the TNG.}}
\subtitle{GIANO-B spectroscopy of red supergiants in Alicante 7 and Alicante 10}

   \author{L. Origlia\inst{1}
          \and E. Dalessandro \inst{1} 
           \and N. Sanna \inst{2} 
          \and A. Mucciarelli \inst{3,1} 
           \and E. Oliva \inst{2}   
        \and G. Cescutti \inst{4}
            \and M. Rainer \inst{2}   
          \and A. Bragaglia \inst{1} 
          \and G. Bono \inst{5,6} 
          }

\institute{
             INAF - Osservatorio di Astrofisica e Scienza dello Spazio di Bologna,
             Via Gobetti 93/3, I-40129, Bologna, Italy
             \email{livia.origlia@inaf.it}
         \and
              INAF - Osservatorio Astrofisico di Arcetri,
              Largo E. Fermi 5, I-50125, Firenze, Italy
         \and
             Dipartimento di Fisica e Astronomia, Universit\'a degli Studi di Bologna,
             Via Gobetti 93/2 - I-40129, Bologna, Italy
         \and
             INAF - Osservatorio Astronomico di Trieste, 
             Via G.B. Tiepolo 11, I-34143, Trieste, Italy
        \and
             Dipartimento di Fisica e Astronomia, Universit\'a degli Studi di Roma Tor Vergata, 
             Via della Ricerca Scientifica 1, I-00133, Roma, Italy
        \and 
            INAF - Osservatorio Astronomico di Roma,
            Via Frascati 33, I-00040, Monte Porzio Catone, Italy
             }

\authorrunning{Origlia et al.}
\titlerunning{GIANO-B spectroscopy of red supergiants in Alicante 7 and Alicante 10}

   \date{Received .... ; accepted ...}

 
  \abstract
   {}
   {The Scutum complex in the inner disk of the Galaxy hosts a number of young clusters and 
associations of red supergiant stars that are heavily obscured by dust extinction.
These stars are important tracers of the recent star formation and chemical enrichment history 
in the inner Galaxy.
   }
   {Within the SPA Large Programme at the TNG, 
we secured GIANO-B high-resolution (R$\simeq$50,000) YJHK spectra 
of 11 red supergiants toward the Alicante 7 and Alicante 10 associations near the RSGC3 cluster. 
   Taking advantage of the full YJHK spectral coverage of GIANO in a single exposure, we were able to 
measure several hundreds of atomic and molecular lines that are suitable for chemical abundance determinations. 
We also measured a prominent diffuse interstellar band at $\lambda$1317.8 nm (vacuum). This provides an independent reddening estimate.
   }
{The radial velocities, Gaia proper motions, and extinction of seven red supergiants in Alicante 7 and three  in Alicante 10 are consistent 
with them being members of the associations.
One star toward Alicante 10
has  kinematics and low extinction that are inconsistent with a membership.
By means of spectral synthesis and line equivalent width measurements,  
we  obtained chemical abundances for iron-peak, CNO, alpha,  
other light, and a few neutron-capture elements.
We found average slightly subsolar iron abundances and solar-scaled [X/Fe] abundance patterns 
for most of the elements, consistent with a thin-disk chemistry.
We found depletion of [C/Fe], enhancement of [N/Fe], and relatively low $\rm ^{12}C/^{13}C<$15, which is
consistent with CN cycled material and possibly some additional mixing in their atmospheres.
}
   {}


   \maketitle
%

%
\section{Introduction}

Stellar  population astrophysics experiences a golden era because of the Gaia mission and ongoing and near-future
massive photometric, astrometric, and spectroscopic surveys
that set the observational framework for an exhaustive
description of the structure of the Milky Way (MW) and its satellites.
Optical and near-infrared (NIR) high-resolution (HR) spectroscopy is crucial to provide the complementary detailed
chemical tagging of selected stellar populations, in order to constrain timescales and overall formation and
chemical enrichment scenarios.
Only echelle spectra with simultaneous HR and wide spectral coverage might be able to allow  measurements of the full set of
iron-peak, CNO, alpha, neutron-capture, Li, Na, Al,  and  other light element abundances with the necessary high accuracy.
These different elements are synthesized in stars and supernovae with different mass progenitors and are therefore released into the interstellar 
medium with different time delays with respect to the onset of the star formation event.

\subsection{SPA Large Programme}

The unique combination of the High Accuracy Radial velocity Planet Searcher for the Northern emisphere (HARPS-N) and GIANO-B echelle spectrographs at the 
Telescopio Nazionale Galileo (TNG), which together cover almost the full optical and NIR range out to the K band,
is ideal to sample the luminous stellar populations of the 
the MW thin disk over almost its
entire extension, as seen from the Northern Hemisphere  at a spectral resolution R$\ge$50,000.
Thus, taking advantage of this HR capability, we proposed a Large Program 
called the {\it SPA - Stellar Population Astrophysics:  detailed age-resolved
chemistry of the Milky Way disk} (Program ID A37TAC\_13, PI: L. Origlia). Observing time has been granted starting from June 2018.
The SPA Large Programme
aims at obtaining high-quality spectra of more than 500 stars in the MW thin disk and associated star
clusters at different Galactocentric distances, including the poorly explored inner disk.
We will observe luminous giant and supergiant stars in young star clusters and associations, luminous Cepheid and Mira variables 
across the entire thin disk, and main-sequence (MS) and other evolved stars of open clusters in the solar neighborhood.

The proposed observations will allow us to provide a detailed mapping of possible gradients, cosmic spreads, and other inhomogeneities 
of individual abundances and abundance ratios. We will also be able to answer a number of open questions
regarding disk formation, evolution, and chemical enrichment, and we will perform critical tests of stellar evolution and nucleosynthesis. This will maximize the scientific return and the overall legacy value for astrophysics.

Based on this detailed chemical tagging and complementary kinematic and evolutionary information from the Gaia mission and other surveys,
we expect in particular to set the framework for a comprehensive chemodynamical modeling of the disk formation and evolution. We hope to be able to distinguish radial migration
from in situ formation scenarios \citep[e.g.,][and references therein]{dasilva16} 
and the overall chemical enrichment history.
In addition, given the expected high quality of the acquired spectra and the target properties, we will be able to perform
the following stellar evolution tests and calibrations: 
i) a first systematic investigation of the dependence of period-luminosity relations of variable stars on their detailed chemical content,
ii) a probe of the poorly explored physics and nucleosynthesis of red supergiants (RSGs), and
iii) a probe of crucial stellar evolution parameters, such as convection,
diffusion, mixing, rotation, and magnetic fields in MS and evolved stars for different stellar ages and evolutionary stages.

\section{High-resolution spectroscopy in the Scutum complex}

\begin{table*}
\caption{RSG stars in the Alicante 7 and Alicante 10 associations of RSGC3 observed with GIANO-B.}
\label{tab1}     
\begin{center}
\begin{tabular}{lllllllcclcc}
\hline\hline
Star &RA(J2000) &DEC(J2000) &J &H &K &T$_{eff}$ &$\xi $ &EW$_{\lambda 1317.8}$ &$A_V$ &RV$_{hel}$ &RV$_{LSR}$\\
\hline
AL07-F1S02  & 18:44:46.9 &  -03:31:07.5  &   8.78  &  6.71  & 5.76  & 3700 & 3.0 & 0.224 &11.6  & 63 & 79\\
AL07-F1S04  & 18:44:39.4 &  -03:30:00.5  &   8.66  &  6.76  & 5.82  & 3600 & 3.0 & 0.211 &10.9  &107 &123\\ 
AL07-F1S05  & 18:44:31.0 &  -03:30:49.9  &   9.87  &  7.93  & 6.97  & 3700 & 3.0 & 0.252 &13.0  & 82 & 98\\ 
AL07-F1S06  & 18:44:29.4 &  -03:30:02.5  &   8.45  &  6.39  & 5.41  & 3600 & 4.0 & 0.247 &12.7  & 79 & 95 \\ 
AL07-F1S07  & 18:44:27.9 &  -03:29:42.7  &   8.06  &  5.91  & 4.85  & 3600 & 4.0 & 0.233 &12.0  & 84 &100\\ 
AL07-F1S08  & 18:44:30.4 &  -03:28:47.1  &   7.34  &  5.20  & 4.10  & 3500 & 4.0 & 0.219 &11.3  & 73 & 89 \\
AL07-F1S09  & 18:44:20.5 &  -03:28:44.7  &   9.28  &  7.24  & 6.21  & 3600 & 3.0 & 0.229 &11.8  & 75 & 91\\
\hline
AL10-C06    & 18:45.17.0 &  -03:41:26.0 &   9.56  &  7.29  & 6.15 & 3600 & 3.0 & 0.252 &13.0  & 75 & 91 \\ 
AL10-C09    & 18:45:11.8 &  -03:40:53.1 &   9.34  &  6.86  & 5.62 & 3600 & 3.0 & 0.299 &15.4  & 76 & 92\\
AL10-C10    & 18:45:11.2 &  -03:39:34.1  &   8.80  &  6.47  & 5.35 & 3600 & 4.0 & 0.242 &12.5  & 74 & 90\\
\hline
AL10-N03    & 18:45:36.6 &  -03:39:19.1  &   7.19  &  5.82  & 5.23 & 3700 & 3.0 & 0.075 &4.0 & 17 & 33\\
\hline
\end{tabular}
\end{center}
{\bf Note}: Identification names, coordinates and 2MASS JHK magnitudes 
from \citet{neg11} for Alicante 7 and \citet{gon12} for Alicante 10.
Effective temperature (T$_{eff})$ in units of Kelvin and microturbulence ($\xi$) in units of km/s from spectral synthesis of GIANO-B spectra.
The EW of the DIB feature at $\lambda$1317.8 nm (vacuum) are in units of nanometers and 
heliocentric RVs in units of km/s, also from the GIANO-B spectra. 
The derived $A_V$ extinction values from the DIB feature at $\lambda$1317.8 nm are based on the calibration with E(B-V) by 
\citet{ham15,DIB_Hamano} and are converted into $A_V$ by using R=3.1.
The LSR RVs were computed using the solar motion from \citet{sch10}.\\
\end{table*}

In the central region of the MW close to the base of the Scutum-Crux arm and the tip of the long bar,
at a distance of $\approx$3.5 kpc from the Galactic center, lies a giant region of star formation that is known as 
the Scutum complex.  This region is characterized by three young and  massive clusters: 
RSGC1 \citep{figer06}, RSGC2  \citep{davies07}, and RSGC3 \citep{clark09,alexander09}. They host a conspicuous population of 
RSG stars with stellar masses 
of $\sim$14-20 $M_{\odot}$ and ages between 12 and 20 Myr. 
A  total 
mass of 2-4$\rm \times 10^{4} M_{\odot}$ has been estimated for these clusters. 
Some extended associations of RSGs  have also been identified around them: 
Alicante 8, in the proximity of RSGC1 and RSGC2 \citep{neg10}, and 
Alicante 7 \citep{neg11} and Alicante 10 \citep{gon12}, which are
associated with RSGC3.

The spectroscopic characterization of the chemical and kinematic properties 
of the stellar poplulations in the Scutum complex started only recently and used IR  spectroscopy because of the huge (A$_V$>10) and patchy extinction that affects 
the Galactic plane in this direction at optical wavelengths.

First radial velocity (RV) measurements suggested 
that these young clusters and associations might share a common 
kinematics. For example, \citet{davies07} and \citet{davies08} 
derived average local standard of rest (LSR) RVs of +109 and +123 km/s for RSGC2 and RSGC1, respectively, 
from K-band spectra with medium-high resolution of the CO bandheads.
\citet{neg11} obtained average  LSR RVs of +102 and +95 km/s for RSGC2 and RSGC3, respectively,
from tmeasurement of the Ca~II triplet lines.

Some chemical abundances of Fe, C, O, and other alpha elements were derived for stars in 
RSGC1 and RSGC2 \citep{davies09b} using NIRSPEC-Keck spectra at a resolution R$\approx$17,000. They
suggest slightly subsolar abundances.

RSGs are very luminous NIR sources even in
highly reddened environments, such as the inner Galaxy, and
they can be spectroscopically studied at HR even
with 4 m class telescopes when these are equipped with
efficient echelle spectrographs. 
A few years ago, during the commissioning and science verification of GIANO
\citep[see, e.g.,][]{oli12a,oli12b,oli13,giano14}, 
we therefore acquired HR (R$\approx$50,000) YJHK spectra of a few RSGs in the RSGC2 and RSGC3 clusters.

In particular, we observed three  RSGs in RSGC2 in July 2012 and five RSGs  RSGC3 in October 2013. 
Detailed RVs and chemical abundances of iron-peak, CNO, alpha, and a few other metals 
have been published in \citet{ori13,ori16}. They
confirmed slightly subsolar iron and iron-peak elements, some depletion of carbon and enhancement of nitrogen, and
approximately solar-scale [X/Fe] abundance ratios for the other metals.

Prompted by the results of these pilot projects, we decided to perform a more systematic 
investigation of the RSGs in the Scutum complex.
As part of the SPA Large Program at the TNG,
we are observing a representative sample of luminous RSG stars, candidate members of the young clusters and associations in the Scutum complex, using the refurbished GIANO-B spectrograph. 
This first paper presents the derived RVs and chemical abundances for RSG 
candidate members of the Alicante 7 and Alicante 10 associations near the RSGC3 star cluster.

\section{Observations and data reduction}

GIANO \citep{giano14} is the HR 
(R$\simeq$50,000) YJHK (950--2450 nm) spectrometer of the TNG 
telescope. GIANO was designed to receive light directly from a dedicated focus of the TNG.
The instrument was 
provisionally commissioned in 2012 and was used in the GIANO-A configuration. In this position,
the spectrometer was positioned on the rotating building and fed through a pair 
of fibers that were connected to another focal station \citep{tozzi14}. The spectrometer was eventually moved to the originally foreseen 
configuration in 2016. This configuration is called GIANO-B \citep{tozzi16}. It can here also be used in the 
GIARPS mode for simultaneous observations with HARPS-N.

GIANO-B provides a fully automated online data reduction pipeline based on the 
GOFIO reduction software \citep{gofio} that processes all the observed 
data, from the calibrations (darks, flats, and U-Ne lamps taken in 
daytime) to the scientific frames. The main feature of the GOFIO data 
reduction is the optimal spectral extraction and wavelength calibration 
based on a physical model of the spectrometer that accurately matches 
instrumental effects such as variable slit tilt and order curvature over 
the echellogram \citep{giano_2Dreduction}. The data reduction package also 
includes bad pixel and cosmic removal, sky and dark subtraction, and 
flat-field and blaze correction. It outputs the scientific data with 
different formats, including merged and non-merged echelle orders.

The GIANO guiding system uses the 850-950 nm light of the target itself
as reference. When reddening is huge, as in the Scutum complex, the limiting factor 
therefore becomes the z-band magnitude of the target.
For observations with GIANO-B we therefore selected those RSGs from the compilations of \citet{neg11} and \citet{gon12}
with typically J$<$10 and A$_V<$16, that is bright enough for both the spectrograph and the guiding camera.

Seven RSGs toward Alicante 7 and four RSGs toward Alicante 10 
(see Table~\ref{tab1}) have been observed with GIANO-B on June-July 2018. 
For the best subtraction 
of the detector artifacts and background, the spectra were collected 
by nodding the star along the slit, that is, with the target alternatively 
positioned at 1/4 (position A) and 3/4 (position B) of the slit length. 
Integration time was 5 minutes per A,B position. The nodding sequences 
were repeated to achieve a total integration time of 40 
minutes per target.

The spectra were reduced using the offline version of GOFIO (available 
at \url{https://atreides.tng.iac.es/monica.rainer/gofio}).

The telluric absorption features were corrected using the spectra of a 
telluric standard (O-type star) taken at different airmasses during the 
same nights. The normalized spectra of the telluric standard taken at 
low and high airmass values were combined with different weights to 
match the depth of the telluric lines in the stellar spectra.

\section{Spectral analysis}

Accurate (better than 1 km/s) RVs and chemical abundances for the observed RSGs 
have been obtained by comparing the 1D GIANO-B spectra with suitable 
synthetic templates. This was previously done in \citet{ori13,ori16} for RSGs in RSGC2 and RSGC3.

In order to compute the required grid of synthetic spectra to model the observed RSGs, 
we used an updated version \citep{ori02} of the code
that was first described in \citet{ori93}. 
The code uses the LTE approximation and the MARCS model atmospheres
\citep{gus08}.
Thousands of NIR atomic transitions from the Kurucz
database\footnote{http://www.cfa.harvard.edu/amp/ampdata/kurucz23/sekur.html},
from \citet{bie73} and \citet{mel99}, and molecular data from
our \citep[][ and subsequent updates]{ori93} compilation and from the compilation of B. Plez
(private communication) are included.  
From this database, a comprehensive list of suitable lines for each
chemical element that can be measured in the RSG spectra, free from significant blending and without strong wings, was extracted \citep[see also][]{ori13} and used to compute equivalent widths.
We used the \citet{gre98} abundances for the solar reference.  

Hyperfine structure splitting was accounted for in the Ni, Sc, and Cu line profile computations. 
Including them does not significantly affect the abundance estimates, however.

\subsection{Stellar parameters}

\begin{figure}
\begin{center}
\includegraphics[width=\hsize]{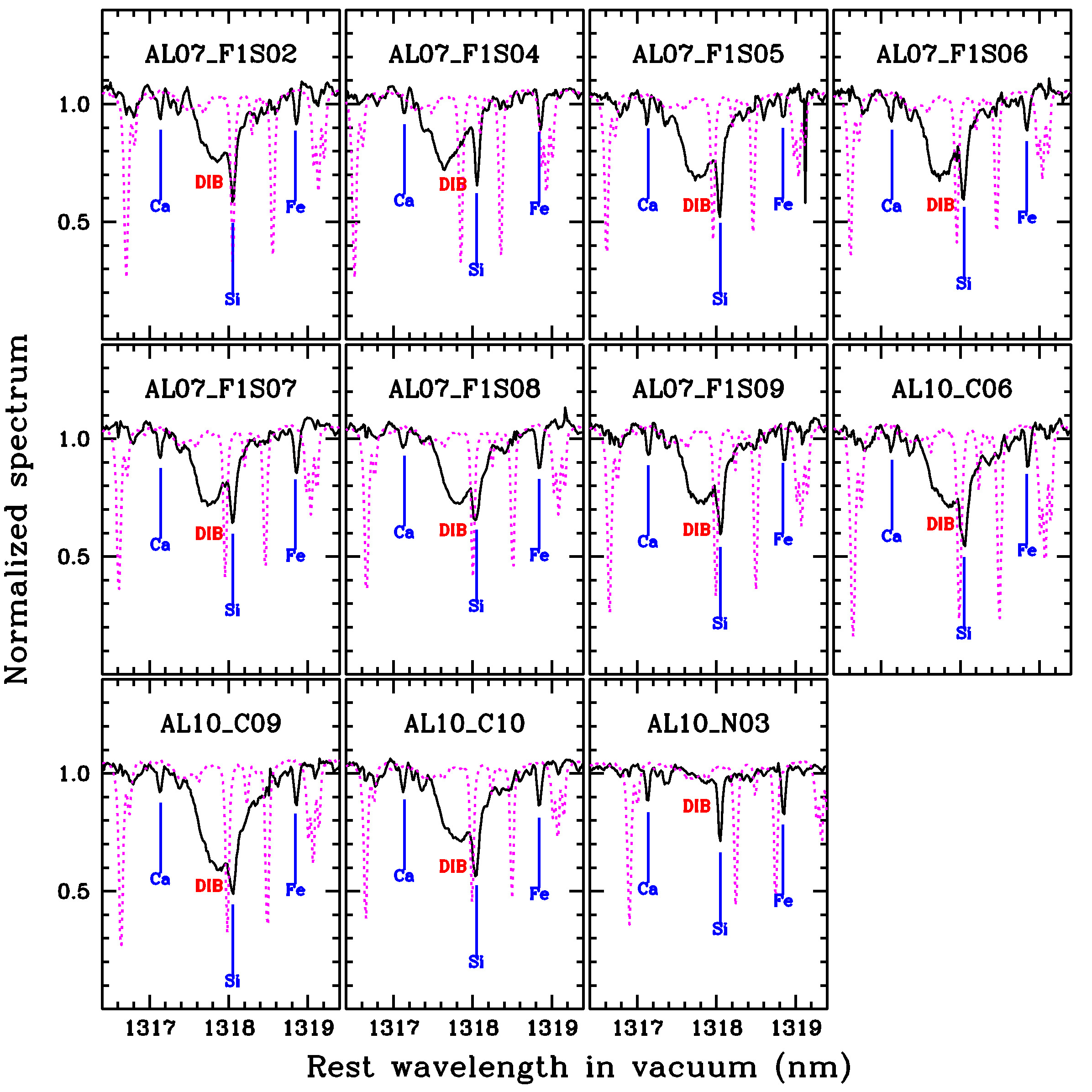}
\end{center}
\caption{Portion of the normalized telluric corrected
spectra, showing the prominent DIB feature
at 1317.8 nm. The dashed purple line is the telluric correction applied
to the data.}
\label{figure_DIB}
\end{figure}

In RSG stars, the strength of the OH and CN molecular lines is especially sensitive to temperature variations.
Moreover,  CN lines are also sensitive to gravity, and OH lines are sensitive to microturbulence.
While the shape and broadening of the  $^{12}$CO bandheads mostly depend on microturbulence, their strengths 
is a quite sensitive thermometer in the 3800-4500~K range, 
where temperature sets the fraction of molecular versus atomic carbon. 
At temperatures below 3800 K most of the carbon is in molecular
form, which drastically reduces the dependence of the CO band
strengths on the temperature itself. At temperatures $\ge$4500~K,
CO molecules barely survive, most of the carbon is in atomic form,
and the CO spectral features become very weak.
A simultaneous reasonable fit of all the molecular lines and bandheads can be obtained only 
in a narrow range of temperature, gravity, and microturbulence, regardless of the adopted CNO abundances.

Moreover, although significantly different temperatures can be inferred
using different scales \citep{lev05, dav13}, synthetic spectra with 
temperatures that are significantly different (especially significantly warmer) than those providing the best-fit solutions
can barely fit the observed spectra and require very peculiar (unlikely) CNO abundance patterns 
\citep[see also][]{ori13,ori16}.

As has been discussed in \citet{ori13,ori16},
at the GIANO spectral resolution of 50,000, we find that in RSGs variations of $\pm$100K in T$_{eff}$, $\pm$0.5 dex in log~g, and $\pm$0.5 km/s 
in microturbulence have effects on the spectral lines that can also be appreciated by some visual inspection.
Variations in  temperature, gravity, and microturbulence smaller than the above values are difficult to distinguish 
because the sensitivity of the lines is limited and the stellar parameters are degenerate. 
Moreover, such a finer tuning would have negligible effect on the inferred abundances (less than a few hundredths of a dex, i.e., smaller than the measurement errors). 
The effect of using different assumptions for the stellar parameters on the derived 
abundances is discussed in Sect.~\ref{abun}.

\subsection{Chemical abundances}
\label{abun}
Chemical abundances were derived by minimizing the scatter between observed and synthetic spectra 
with suitable stellar parameters, by using spectral synthesis and, as figures of merit, line equivalent widths 
and statistical tests on both the difference between the model and the observed spectrum and on the $\chi^2$.
As discussed in \citet{ori13} and references therein, 
the simple difference is
more powerful for quantifying systematic discrepancies than the classical $\chi ^2$ test, 
which is instead equally sensitive to {\em \textup{random}} and {\em \textup{systematic}} deviations.

The typical random error of the
measured line equivalent widths is $<$10 m\AA. It mostly arises from a
$\pm $1-2\% uncertainty in the placement of the pseudo-continuum, as
estimated by overlapping the synthetic and observed spectra.  This
error corresponds to abundance variations of $<$0.1 dex, that is, smaller
than the typical 1$\sigma $ scatter ($<$0.15 dex) in the derived
abundances from different lines.  The errors quoted in Table~\ref{tab2}
for the final abundances were obtained by dividing the 1$\sigma $
scatter by the square root of the number of lines we used, typically from a
few to a few tens per species.  When only one line was available, we assumed an error of
0.10~dex.

As detailed in \citet{ori13}, a somewhat conservative estimate of the overall systematic uncertainty in the abundance 
(X) determination, caused by 
variations of the atmospheric parameters, can be computed with the following formula:
$\rm (\Delta X)^2 = (\partial X/\partial T)^2 (\Delta T)^2 + (\partial X/\partial log~g)^2 (\Delta log~g)^2  + (\partial X/\partial \xi)^2 (\Delta \xi)^2$.
We computed test models with variations of  
$\pm $100~K in temperature, $\pm$0.5~dex in log~g, and $\pm $0.5 km/s in microturbulence velocity 
with respect to the best-fit parameters. We found that these systematic uncertainties in stellar parameters 
can affect the overall abundances at a level of 0.15-0.20 dex. 

\subsection{Diffuse interstellar bands: extinction and contamination}

  \begin{figure*}
  \centering
  \includegraphics[width=\hsize]{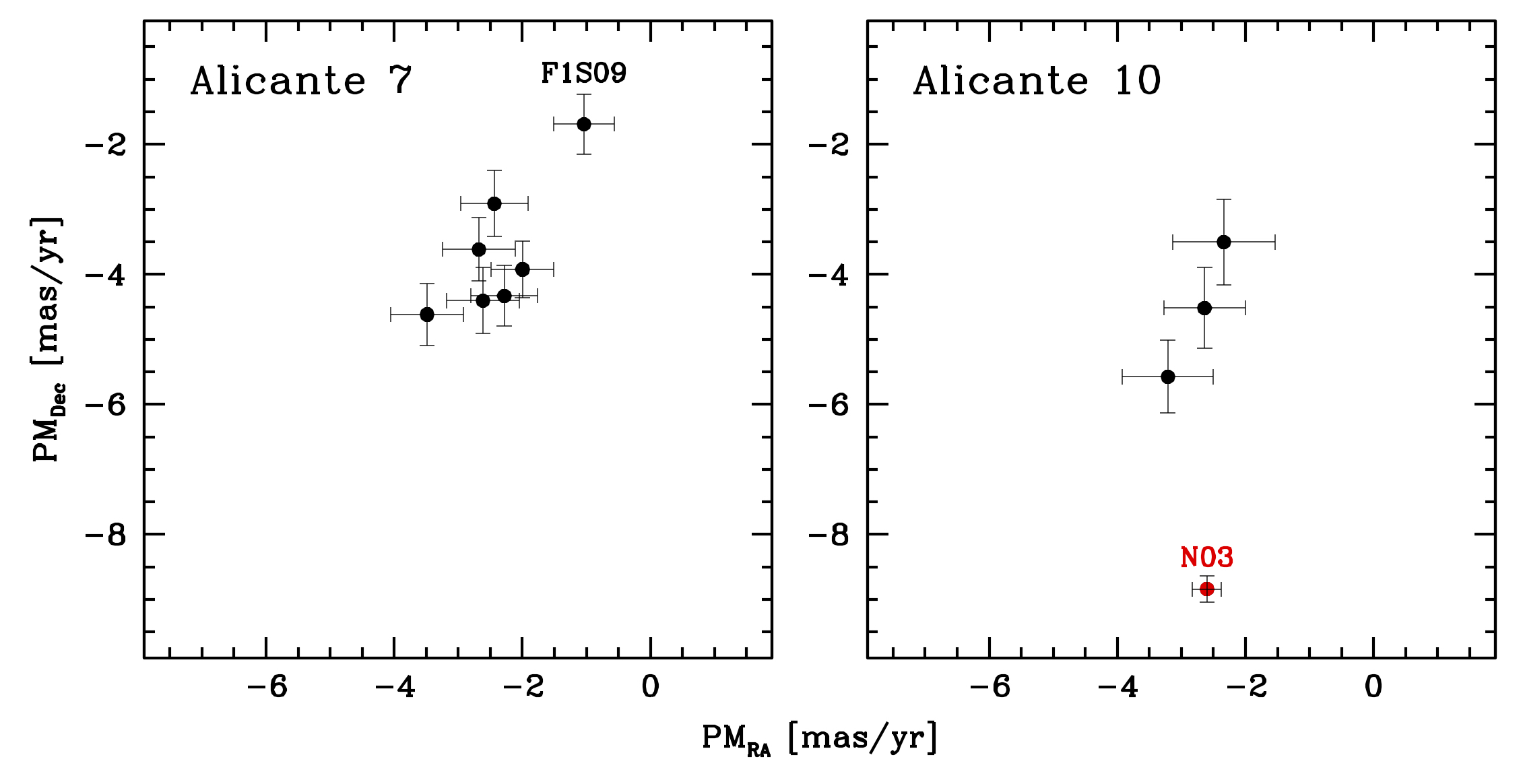}
   \caption{Gaia DR2 proper motions for the observed stars in Alicante 7 (left panel) and Alicante 10 (right panel).
       The red dot indicates star N03, whose kinematics is inconsistent with being a member of the Alicante 10 association.}
              \label{figpm}
    \end{figure*}

\begin{table*}
\footnotesize
\caption{Chemical abundances ([X/H]) of the RSG stars in the Alicante 7 and Alicante 10 associations observed with GIANO-B.}
\label{tab2}     
\tabcolsep 3pt
\begin{tabular}{lllllllllllll}  
\hline\hline
Element$^a$&\multicolumn{7}{c}{Alicante 7$^b$}&\vline& \multicolumn{4}{c}{Alicante 10$^b$} \\ 
\hline
&F1S02&F1S04&F1S05&F1S06&F1S07&F1S08&F1S09&\vline&C06&C09&C10&N03\\
\hline
Fe(26)&-0.18(23)&-0.12(17)&-0.13(17)&-0.23(30)&-0.21(23)&-0.28(30)&-0.12(24)&\vline&-0.18(15)&-0.23(20)&-0.28(24)&-0.17(21)\\
 &$\pm $0.04&$\pm $0.05&$\pm $0.05&$\pm $0.04&$\pm $0.04&$\pm $0.04&$\pm $0.05&\vline&$\pm $0.05&$\pm $0.05&$\pm $0.05&$\pm $0.04\\             
\hline
C(6)&-0.46(8)&-0.45(7)&-0.50(9)&-0.52(9)&-0.43(9)&-0.49(9)&-0.42(8)&\vline&-0.46(7)&-0.49(9)&-0.45(8)&-0.53(9)\\
 &$\pm $0.07&$\pm $0.04&$\pm $0.06&$\pm $0.05&$\pm $0.05&$\pm $0.04&$\pm $0.05&\vline&$\pm $0.07&$\pm $0.07&$\pm $0.04&$\pm $0.05\\             
N(7)& 0.18 (48)& 0.25(44)& 0.27(16)& 0.20(51)& 0.06(74)& 0.07(61)& 0.27(32)&\vline& 0.19(37)& 0.16(50& 0.13(67)& 0.04(72)\\
 &$\pm $0.03&$\pm $0.03&$\pm $0.04&$\pm $0.03&$\pm $0.02&$\pm $0.03&$\pm $0.04&\vline&$\pm $0.03&$\pm $0.03&$\pm $0.03&$\pm $0.02\\             
O(8)&-0.20 (19)&-0.12(16)&-0.20(15)&-0.25(17)&-0.24(19)&-0.28(11)&-0.15(15)&\vline&-0.24(16)&-0.19(16)&-0.27(17)&-0.22(11)\\
 &$\pm $0.04&$\pm $0.03&$\pm $0.05&$\pm $0.04&$\pm $0.05&$\pm $0.06&$\pm $0.04&\vline&$\pm $0.04&$\pm $0.06&$\pm $0.04&$\pm $0.05\\             
F$^c$(9)&-0.24 (1)&-0.14(1)&-0.29(1)&-0.20(1)&-0.23(1)&-0.33(1)&-0.11(1)&\vline&-0.17(1)&-0.09(1)&-0.22(1)&-0.11(1)\\
 &$\pm $0.10&$\pm $0.10&$\pm $0.10&$\pm $0.10&$\pm $0.10&$\pm $0.10&$\pm $0.10&\vline&$\pm $0.10&$\pm $0.10&$\pm $0.10&$\pm $0.10\\             
Na(11)&-0.07(1)&-0.11(1)&-0.15(1)&-0.14(1)&-0.17(1)&-0.24(1)&-0.17(1)&\vline&-0.15(1)&-0.14(1)&-0.06(2)&-0.11(1)\\
 &$\pm $0.10&$\pm $0.10&$\pm $0.10&$\pm $0.10&$\pm $0.10&$\pm $0.10&$\pm $0.10&\vline&$\pm $0.10&$\pm $0.10&$\pm $0.16&$\pm $0.10\\             
Mg(12)&-0.18(6)&-0.08(3)&-0.10(3)&-0.18(5)&-0.17(7)&-0.27(8)&-0.11(3)&\vline&-0.12(3)&-0.20(4)&-0.26(3)&-0.17(3)\\
 &$\pm $0.06&$\pm $0.13&$\pm $0.09&$\pm $0.11&$\pm $0.11&$\pm $0.06&$\pm $0.12&\vline&$\pm $0.11&$\pm $0.15&$\pm $0.04&$\pm $0.03\\             
Al(13)&-0.14(4)&-0.19(4)&-0.14(3)&-0.23(5)&-0.23(5)&-0.32(4)&-0.18(3)&\vline&-0.15(4)&-0.25(4)&-0.22(4)&-0.19(4)\\
 &$\pm $0.07&$\pm $0.02&$\pm $0.06&$\pm $0.04&$\pm $0.05&$\pm $0.04&$\pm $0.02&\vline&$\pm $0.02&$\pm $0.04&$\pm $0.02&$\pm $0.03\\             
Si(14)&-0.17(14)&-0.13(11)&-0.09(4)&-0.18(14)&-0.20(19)&-0.22(21)&-0.11(10)&\vline&-0.10(5)&-0.20(12)&-0.18(10)&-0.20(13)\\
 &$\pm $0.05&$\pm $0.07&$\pm $0.11&$\pm $0.05&$\pm $0.04&$\pm $0.04&$\pm $0.06&\vline&$\pm $0.08&$\pm $0.04&$\pm $0.04&$\pm $0.04\\             
P(15)&-0.10(2)&-0.16(2)&-0.17(1)&-0.09(2)&-0.20(1)&-0.22(1)&-0.18(1)&\vline&-0.17(1)&-0.07(1)&-0.14(2)&-0.19(2)\\
 &$\pm $0.07&$\pm $0.08&$\pm $0.10&$\pm $0.07&$\pm $0.10&$\pm $0.10&$\pm $0.10&\vline&$\pm $0.10&$\pm $0.10&$\pm $0.02&$\pm $0.02\\                             
S(16)&-0.11(1)&-0.11(1)&-0.08(1)&-0.18(1)&-0.29(1)&-0.24(1)&-0.12(1)&\vline&-0.15(1)&-0.26(1)&-0.28(1)&-0.16(1)\\
 &$\pm $0.10&$\pm $0.10&$\pm $0.10&$\pm $0.10&$\pm $0.10&$\pm $0.10&$\pm $0.10&\vline&$\pm $0.10&$\pm $0.10&$\pm $0.10&$\pm $0.10\\
K(19)&-0.03(1)&-0.05(2)&-0.10(2)&-0.05(2)&-0.20(2)&-0.17(2)&-0.08(1)&\vline&--      &-0.00(1)&-0.24(2)&--      \\
 &$\pm $0.10&$\pm $0.09&$\pm $0.06&$\pm $0.02&$\pm $0.08&$\pm $0.04&$\pm $0.10&\vline&   &$\pm $0.10&$\pm $0.03&  \\              
Ca(20)&-0.21(7)&-0.17(8)&-0.14(7)&-0.26(8)&-0.26(8)&-0.32(6)&-0.15(7)&\vline&-0.15(5)&-0.21(5)&-0.21(8)&-0.17(5)\\
 &$\pm $0.04&$\pm $0.04&$\pm $0.06&$\pm $0.06&$\pm $0.03&$\pm $0.04&$\pm $0.06&\vline&$\pm $0.09&$\pm $0.09&$\pm $0.06&$\pm $0.04\\             
Sc(21)&-0.15(2)&-0.14(2)&-0.24(2)&-0.27(2)&-0.18(4)&-0.30(4)&-0.10(2)&\vline&-0.16(2)&-0.21(4)&-0.31(4)&-0.11(1)\\
 &$\pm $0.05&$\pm $0.05&$\pm $0.10&$\pm $0.07&$\pm $0.05&$\pm $0.06&$\pm $0.03&\vline&$\pm $0.06&$\pm $0.07&$\pm $0.06&$\pm $0.10\\             
Ti(22)&-0.18(20)&-0.14(17)&-0.14(20)&-0.18(19)&-0.24(24)&-0.24(23)&-0.17(15)&\vline&-0.14(18)&-0.26(18)&-0.26(22)&-0.21(17)\\
 &$\pm $0.05&$\pm $0.05&$\pm $0.05&$\pm $0.05&$\pm $0.05&$\pm $0.04&$\pm $0.06&\vline&$\pm $0.06&$\pm $0.05&$\pm $0.05&$\pm $0.05\\             
V(23)&-0.18(1)&-0.13(1)&-0.18(1)&-0.27(1)&-0.27(1)&-0.30(1)&-0.15(1)&\vline&-0.17(1)&-0.26(1)&-0.25(1)&-0.17(1)\\
 &$\pm $0.10&$\pm $0.10&$\pm $0.10&$\pm $0.10&$\pm $0.10&$\pm $0.10&$\pm $0.10&\vline&$\pm $0.10&$\pm $0.10&$\pm $0.10&$\pm $0.10\\             
Cr(24)&-0.18(2)&-0.15(4)&-0.15(2)&-0.16(4)&-0.20(3)&-0.22(10)&-0.14(2)&\vline&-0.20(2)&-0.27(2)&-0.30(2)&-0.22(2)\\
 &$\pm $0.00&$\pm $0.01&$\pm $0.05&$\pm $0.05&$\pm $0.05&$\pm $0.06&$\pm $0.02&\vline&$\pm $0.00&$\pm $0.00&$\pm $0.03&$\pm $0.09\\             
Co(27)&-0.12(1)&-0.09(1)&-0.04(1)&-0.23(1)&-0.13(1)&-0.29(1)&-0.08(1)&\vline&-0.11(1)&-0.21(1)&-0.23(1)&-0.22(1)\\
 &$\pm $0.10&$\pm $0.10&$\pm $0.10&$\pm $0.10&$\pm $0.10&$\pm $0.10&$\pm $0.10&\vline&$\pm $0.10&$\pm $0.10&$\pm $0.10&$\pm $0.10\\             
Ni(28)&-0.21(2)&-0.13(4)&-0.14(3)&-0.24(2)&-0.18(2)&-0.27(3)&-0.19(3)&\vline&-0.20(3)&-0.28(2)&-0.26(2)&-0.20(3)\\
 &$\pm $0.02&$\pm $0.11&$\pm $0.04&$\pm $0.00&$\pm $0.05&$\pm $0.07&$\pm $0.05&\vline&$\pm $0.11&$\pm $0.02&$\pm $0.00&$\pm $0.05\\             
Cu(29)&-0.19(2)&-0.18(2)&-0.17(2)&-0.32(2)&-0.26(2)&-0.27(2)&-0.19(2)&\vline&-0.22(2)&-0.26(2)&-0.26(2)&-0.24(2)\\
 &$\pm $0.03&$\pm $0.06&$\pm $0.05&$\pm $0.06&$\pm $0.04&$\pm $0.03&$\pm $0.05&\vline&$\pm $0.12&$\pm $0.02&$\pm $0.09&$\pm $0.01\\             
Y(39)&-0.29(5)&-0.15(4)&-0.28(3)&-0.30(3)&-0.22(4)&-0.33(5)&-0.16(4)&\vline&-0.15(4)&-0.23(5)&-0.26(4)&-0.19(5)\\
 &$\pm $0.04&$\pm $0.03&$\pm $0.01&$\pm $0.05&$\pm $0.05&$\pm $0.03&$\pm $0.04&\vline&$\pm $0.03&$\pm $0.04&$\pm $0.02&$\pm $0.04\\             
Zr(40)&-0.18(2)&-0.14(1)&-0.11(1)&-0.40(1)&-- &--      &-0.17(2)&\vline&-0.20(1)&-0.31(1)&--     &-0.18(2)\\  
 &$\pm $0.01&$\pm $0.10&$\pm $0.10&$\pm $0.10& &  &$\pm $0.24&\vline&$\pm $0.10&$\pm $0.10&  &$\pm $0.10\\        
\hline
$\rm ^{12}C/^{13}C$~$^d$ & 10& 13& 13& 11& 13& 11& 13&\vline&12& 12& 11& 9 \\
\hline\hline
\end{tabular}

\vspace{0.2cm}
Notes:\\
{\bf $^a$} Chemical element and corresponding atomic number in parentheses.

{\bf $^b$} Numbers in parentheses refer to the number of lines used to compute the abundances.

{\bf $^c$} F abundances are obtained from the HF(1-0) R9 line using the parameters listed in \citet{jon14}. 

{\bf $^d$} Typical error in the $\rm ^{12}C/^{13}C$ isotopic ratio is $\pm$1.

\end{table*}

The observed stars are affected by a large and differential foreground interstellar absorption. 
Therefore, the diffuse interstellar band (DIB) absorption features
\citep{DIB_Geballe,DIB_Cox,DIB_Hamano} 
are expected to be visible in our spectra.  These 
broad features may be conveniently used as an independent estimate of 
the foreground extinction within the limits related to the quite
large scatter of the feature strength 
for stars with similar extinction.

The most prominent DIB feature among those reported in the above references
falls at $\lambda$1317.8 nm. At this wavelength the RSG spectra are remarkably 
free of photospheric lines. This
DIB feature can therefore easily be recognized and accurately measured in our 
spectra (see Figure~\ref{figure_DIB}). 
The measured equivalent widths
are reported in Table~\ref{tab1}, together with the corresponding
estimate of visual extinction ($A_V$), based on the calibration 
with E(B-V) reported in \citet{ham15,DIB_Hamano}. 
We find $A_V$ values in the 10-16 mag range, which is fully consistent with the corresponding $E(J-K)$ values quoted by
\citet{neg11} and \citet{gon12}.

The other DIB features that can be directly measured in our
spectra are $\lambda$1078.3, $\lambda$1262.6 and $\lambda$1280.2. They
are much weaker, and their relative strengths are compatible with
those reported in \cite{DIB_Hamano}.

The other DIB features are much more difficult to recognize in our
spectra because they fall at wavelengths where the spectra are
crowded with photospheric lines of atomic and molecular species. 
To measure these lines would require a simultaneous modeling 
and fitting of the stellar and DIB spectra. This task is well beyond the aims of
this paper.  
Nonetheless, the photospheric lines close
to the DIB features should be used carefully. 
In particular, the region around
the strong DIB feature at $\lambda$1527.3 nm should be avoided, whose equivalent 
width is expected to be similar to that of the feature at $\lambda$1317.8
\citep{DIB_Cox}.

\section{Results}

The best-fit values of the stellar parameters and RVs for the observed RSGs in Alicante 7 and Alicante 10 are 
reported in Table~\ref{tab1}.
We find temperatures in the 3500-3700~K range and microturbulence of 3-4 km/s. 

For all the observed stars we adopted a surface gravity of log~g=0.0. 
For the RSGs in RSGC2 and RSGC3 \citep{ori13,ori16}, the line profiles
of the observed stars in Alicante 7 and Alicante 10 are also definitely broader than the instrumental line profile
(as determined from telluric lines and from the GIANO-B spectra of standard stars).
This additional broadening is likely due to macroturbulence and is normally modeled with a
Gaussian profile, as for the instrumental broadening.
All the analyzed RSGs show macroturbulence velocity of 9-10 km/s,
which is well within the range of values measured in the RSGs
of the Galactic center \citep[see, e.g.][]{ram00,cun07,davies09a}. 
We did not find other appreciable line broadening by stellar rotation.

The derived values of $A_V$ from the DIB feature at 1317.8 nm are consistent with the high 
extinction toward the Scutum complex for all but star AL10-N03 
(see Table~\ref{tab1}), which has a much weaker DIB feature and likely much lower extinction.
Notably, this star also has a significantly lower RV than the other stars in Alicante 10.

The RVs of all the other stars measured in Alicante 7 and Alicante 10 are consistent with them being members of the 
Scutum complex, although they 
show lower values on average than the average value of RV$_{\rm hel}$=90 km/s with a dispersion of 2.3 km/s that is measured in the parent cluster RSGC3 \citep{ori16}.
Only the heliocentric RV of star AL07-F1S04 exceeds 100 km/s.

  \begin{figure}
  \centering
  \includegraphics[width=\hsize]{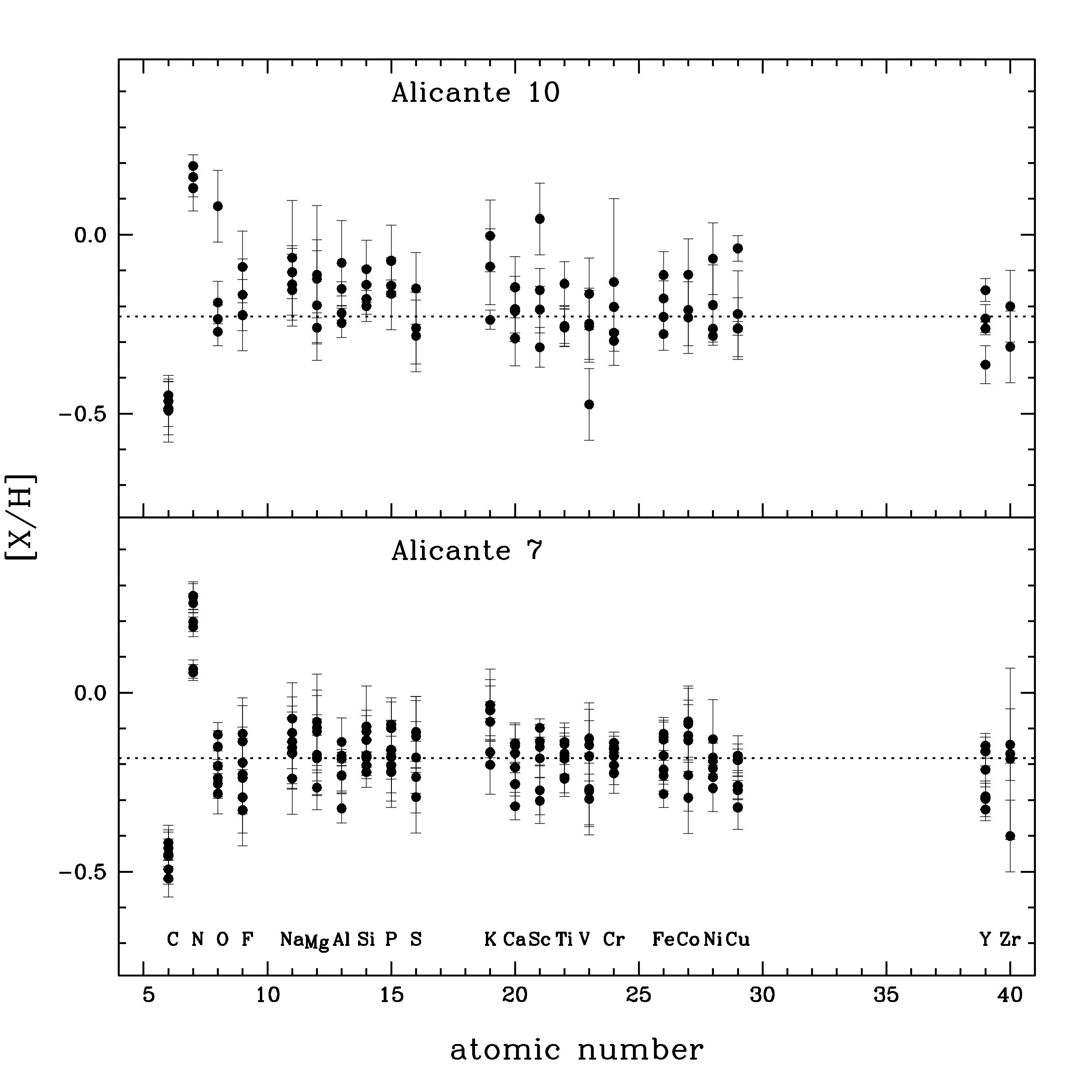}
   \caption{Derived [X/H] chemical abundances and corresponding errors 
   for the observed RSGs candidate members of the Alicante 7 (seven stars) and Alicante 10 (three stars) associations. 
   Dotted lines mark the average [Fe/H] values in each association.}
              \label{figabun}
    \end{figure}

  \begin{figure}
  \centering
  \includegraphics[width=\hsize]{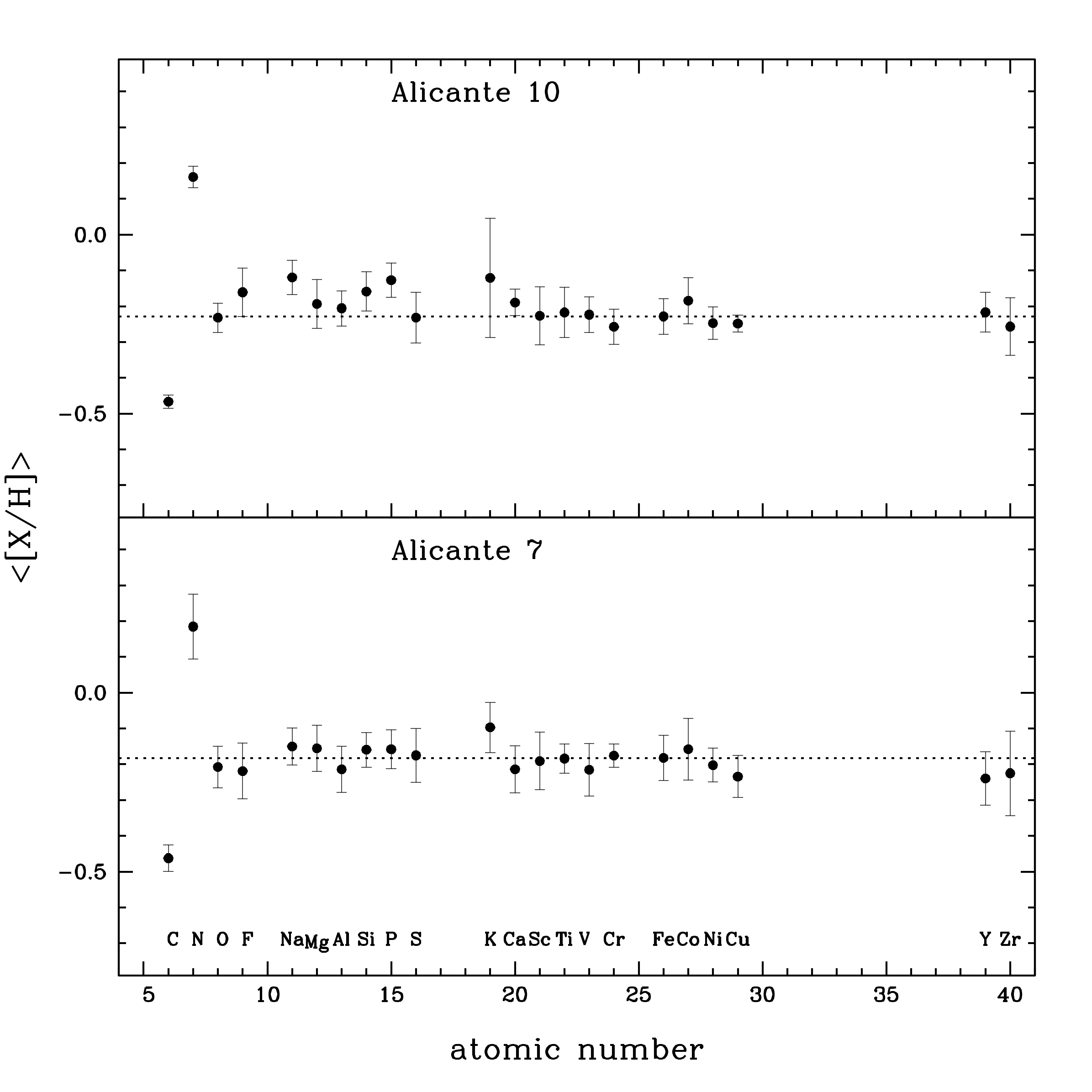}
   \caption{Derived average [X/H] chemical abundances and corresponding errors 
   for the observed RSGs, candidate members of the Alicante 7 and Alicante 10 associations. 
   Dotted lines mark the average [Fe/H] values in each association.}
              \label{meanabun}
    \end{figure}

  \begin{figure}
  \centering
  \includegraphics[width=\hsize]{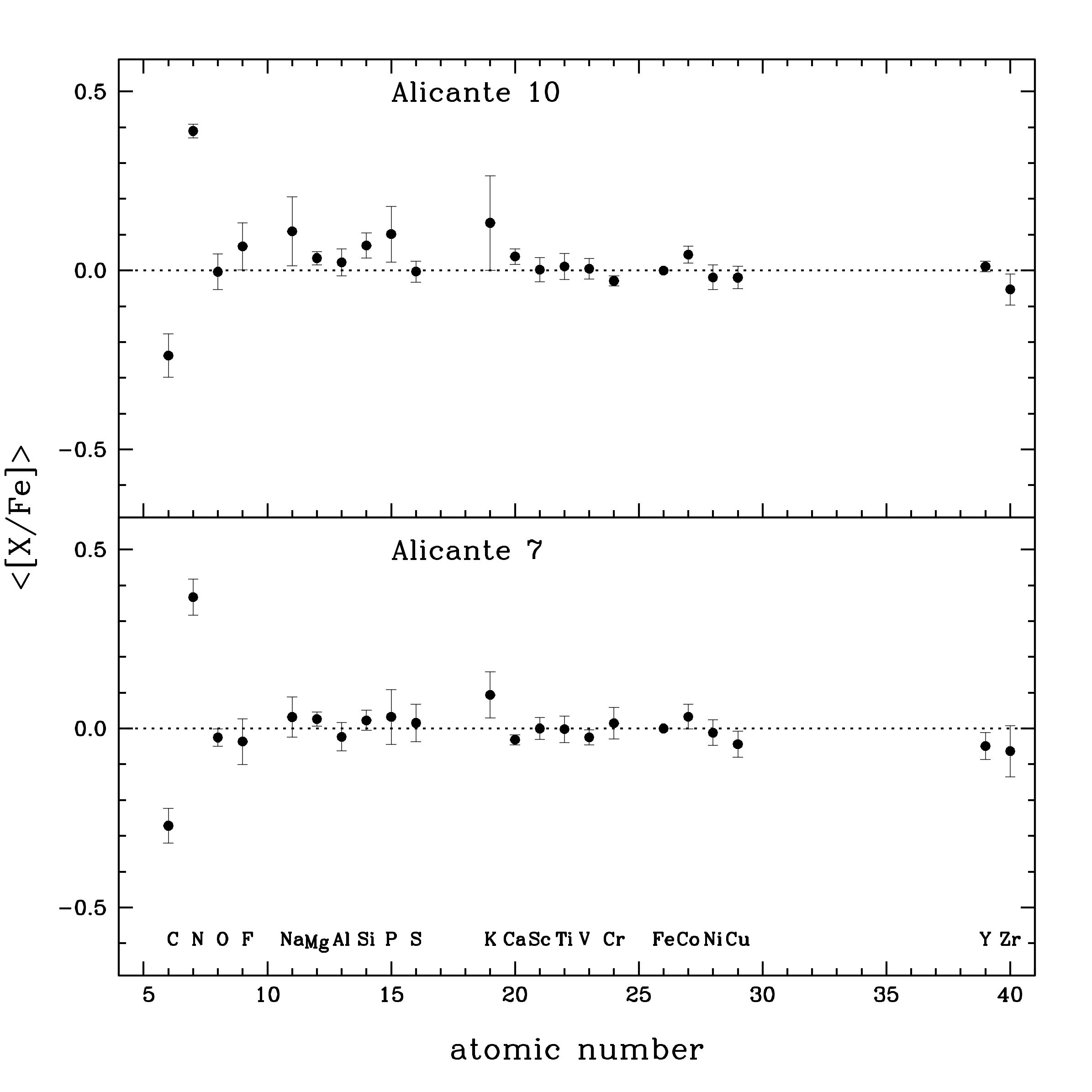}
   \caption{Derived average [X/H] chemical abundances and corresponding errors 
   for the observed RSGs, 
   candidate members of the Alicante 7 and Alicante 10 associations.}
              \label{meanratio}
    \end{figure}

Proper motions for all the observed RSGs in Alicante 7 and Alicante 10 were measured from data of Gaia DR2 \citep{gaia} 
and are plotted in Fig.~\ref{figpm}. 
The RVs and PMs of the RSGs in Alicante 7 are consistent with them being members of the association within the errors. The only possible exception is F1S09, whose proper motions are barely consistent with a membership at $\approx3-4\sigma$ level, 
while its RV is well consistent.
In Alicante 10 the three stars whose RVs are consistent with membership also have consistent proper motions, while the RV of
N03 clearly is too low, the proper motions are too different, and the extinction is too low for it to likely be member of the association. 
In the following we only consider as member RSGs of the Alicante 10 association stars C06, C09, and C10. 

For all the observed stars, Table~\ref{tab2} lists
the derived chemical abundances and their associated errors of 
Fe and other iron-peak elements (V, Cr, Co, Ni, and Cu), CNO, and the other alpha elements (Mg, Si, Ca, and Ti), 
other light (F, Na, Al, P, S, K, and Sc), and a few neutron-capture (Y, Zr) elements.
The atomic lines of V and Cu are heavily blended, therefore their derived abundances should be taken with caution.
The few measurable Mn and SrII lines in the GIANO-B spectra are too strong or saturated to derive reliable abundances. 
While for most of the metals abundances were derived from the measurements of atomic lines, 
for the CNO elements they were obtained from the many molecular lines of CO, OH, and CN in the NIR spectra.
$^{12}$C and $^{13}$C carbon abundances were determined both from a few individual  roto-vibrational lines as well as from bandheads, 
because of the high level of crowding and blending of the CO lines in these stars. We found fully consistent results with the two methods.
The fluorine abundance was determined from the HF(1-0) R9 line using the revised log~$gf$ and excitation potential values 
of \citet{jon14}. A few other HF lines are present in the K-band spectrum of RSGs, but they were to too strongly blended or too faint 
to provide reliable abundance estimates.

Figure~\ref{figabun} shows the chemical abundances of the individual RSGs that are likely members of the 
Alicante 7 (seven stars) and Alicante 10 (three stars) associations. Figures~\ref{meanabun} and~\ref{meanratio} show the mean [X/H] abundances and [X/Fe] abundance ratios for 
all the sampled chemical elements, as computed by averaging the corresponding values for the seven RSGs in Alicante 7 and the 
three RSGs in Alicante 10 that are likely members of these associations, according to the measured RVs and high extinction.

Average values of [Fe/H]=-0.18 and -0.215 dex and corresponding 1$\sigma$ scatter of 0.06 and 0.05~dex for the RSGs that are likely members 
of Alicante 7 and Alicante 10, respectively, were measured. 
All the other chemical elements show about solar-scaled [X/Fe] abundance ratios within $\pm$0.1 dex, with the only exception 
of C and N, which show a depletion of a few tenths of a dex for carbon and and enhancement for nitrogen. The 
$\rm ^{12}C/^{13}C$ isotopic ratios between 9 and 13 were also measured.

\section{Discussion and conclusions}

For the seven RSGs in Alicante 7 and for three of the four RSGs of Alicante 10, the measured LSR RVs in the 79-123 km/s range 
suggest a Galactocentric radius of $\approx$4$\pm0.5$ kpc 
and kinematic near \citep[see, e.g.,][]{rom09} distances in the 5-7 kpc range, suggesting that they are likely located beyond the molecular ring 
and likely members of the Scutum complex.
This is also consistent with the corresponding high extinction measured in their direction (see Table~\ref{tab1}).
For N03 toward Alicante 10 with an RV$_{LSR}$ of 33 km/s,    
Galactocentric radii of $\approx 6$ kpc and  
a near distance of $\approx2.3$ kpc can be derived. This suggests that  
N03 might be located  in front of the molecular ring, which also agrees with its significantly lower extinction 
when compared to the values inferred in the Scutum complex. 

All the metals measured in the RSGs of Alicante 7 and Alicante 10 show slightly subsolar abundances and about solar-scale abundance ratios.
The only remarkable exceptions are the depletion of C, the enhancement of N, and the relatively low (in the 9-13 range) $\rm ^{12}C/^{13}C$ isotopic ratio, 
which are consistent with CN burning and other possible additional mixing processes that occur in the post-MS stages of stellar evolution.
It is worth mentioning that the inferred [C/N] abundance ratios in the $\rm -0.77\ge[C/N]\le-0.49$ range
are also consistent with the surface values predicted by stellar evolution calculations
for RSGs at the end of their lives \citep[e.g.,][]{dav19}.
The rather homogeneous chemistry of Alicante 7 and Alicante 10 is also 
fully consistent with the abundances measured in a few RSGs of the parent RSGC3 cluster \citep{ori16}. 
A more comprehensive chemical and kinematic study of the latter and the overall link with its associations 
will be presented in a forthcoming paper. 

As discussed in \citet{ori16},
the slightly subsolar metallicity measured in regions of recent star formation in the Scutum complex is intriguing because  metal abundances well in excess of solar have been measured in the thin disk at larger Galactocentric distances 
\citep[see, e.g.,][and references therein]{gen13} and were also predicted by the inside-out formation scenario for  the Galactic disk
\citep{cesc07}.
Interestingly enough, the likely foreground RSG AL10-N03 also has slightly subsolar metallicity as well as pre-MS clusters at Galactocentric distances 
from 6.7 to 8.7 kpc that were studied by \citet{spi17}. The innermost of these clusters are 0.10-0.15 dex less metallic than the majority of the older clusters that are located at similar Galactocentric radii. 

This may suggest that recent star formation in the inner disk may have occurred from a gas that was not as enriched as expected. One possible explanation is  dilution by metal-poor halo gas driven there by dynamical interactions of the disk with 
other galactic substructures such as bars, rings, and spiral arms.

A precise description of how this dynamical process could work is beyond the scope of this paper,
but our observations have indeed pointed out another complexity in the evolution 
and chemical enrichment of the inner Galaxy that is to be addressed by future studies.
At the completion of the SPA Large Programme, when a more comprehensive chemical and kinematic mapping of the Scutum complex 
and possibly of other regions of recent star formation will become available, this hypothesis can be verified on 
more comprehensive grounds.

\begin{acknowledgements}
We thank C. Evans, the referee of our paper, for his useful comments and suggestions.
\end{acknowledgements}

\end{document}